\documentstyle[twocolumn,aps,epsf,floats]{revtex} 
\tighten
\begin{document}\draft
\title{Dephasing times in quantum dots due to elastic 
LO phonon-carrier collisions}
\author{A.~V.~Uskov,$^{1,2,4}$  A.-P.~Jauho,$^3$ B.~Tromborg,$^2$ J.~M{\o}rk,$^2$
and R.~Lang$^4$}
\address{
$^1$Lebedev Physical Institute, Leninsky pr. 53, 117924 Moscow, Russia\\
$^2$Research Center COM, Technical University
of Denmark, Bldg. 349, DK-2800 Lyngby, Denmark\\
$^3$Mikroelektronik Centret, Technical University of Denmark, Bldg. 345east,
DK-2800 Lyngby, Denmark\\
$^4$Dept. of Applied Physics, Tokyo University of Agriculture and Technology,
Koganei-shi 184-8588, Japan,\\
and CREST, Japan Science and Technology Corporation
\\\medskip
\date{\today}
\parbox{14cm}{\rm
Interpretation of experiments on 
quantum dot (QD) lasers presents a challenge: the phonon bottleneck, which
should strongly suppress relaxation and dephasing of the discrete
energy states, often seems to be inoperative.
We suggest  and develop a theory for an
intrinsic mechanism for dephasing in QD's:
second-order elastic interaction between quantum
dot charge carriers and LO--phonons. The calculated
dephasing
times are of the order of 200 fs at room temperature, consistent with 
experiments. The phonon bottleneck thus does not prevent
significant room temperature
dephasing.}
\smallskip\\}
\maketitle
\narrowtext

Semiconductor Quantum Dots (QDs) may pave the way
to
optoelectronic devices  with  vastly superior
performance compared to present devices [\onlinecite{Bimberg}].
A detailed understanding of QD's 
optical properties  is therefore of utmost importance, both from
a basic science and from a technological point of view.
The homogeneous linewidth $\Gamma$ of an optical transition,
and the dephasing time, or, equivalently, the polarization relaxation time,
$T_2=2/\Gamma$, are basic characteristics for interaction
of light with quantum systems.
In lasers, the homogeneous linewidth 
can strongly effect the laser gain.  
The dephasing time   defines the time scale on which coherent 
interaction of light with medium takes 
place [\onlinecite{Meystre}], and, in particular, 
it gives the ultimate 
time scale for realization of coherent control in a quantum 
system (see, for instance, Ref. [\onlinecite{Bonadeo}] and references therein).  

In bulk and Quantum Well (QW) 
semiconductors the dephasing time is usually taken   to 
be the intraband relaxation time [\onlinecite{Chow}].  
At low carrier densities, this time
is set by LO 
phonon-carrier scattering while at  higher carrier densities
carrier-carrier collisions are also important.
The energy spectrum of carriers in QDs is discrete, 
and in this sense 
the interaction processes  in QDs are qualitatively different from  
those in QWs or bulk
materials. 
For example, if the LO phonon energy does not coincide 
with the separation of  energy 
levels in a QD,  LO phonon-carrier interaction can 
not lead to carrier relaxation between these 
levels (this is the so called ``phonon bottleneck'' [\onlinecite{Benisty}]).  
At first sight, then, in the absence of this efficient mechanism
much longer relaxation and dephasing
times could be expected.

Carrier relaxation is a result of
inelastic carrier--phonon and carrier--carrier 
collisions.  In contrast, the dephasing time  
$T_2$ is influenced by both inelastic and elastic 
collisions.   Uskov {\it et al.} [\onlinecite{Uskov}]
demonstrated recently that elastic collisions of two-dimensional (2D) 
carriers with carriers confined in self-assembled QDs can lead to 
substantial collisional broadening 
($T_2 \simeq$ 0.1 -- 1 ps at moderate densities of 2D carriers) 
of QD spectral lines. 
In this work we study
{\it elastic} collisions between QD-carriers and LO-phonons:
these collisions 
can disturb the phase of the carrier wave functions in QDs 
without changing the populations of the carrier energy levels, 
and accordingly QD spectral lines will be 
broadened.  
We will show that these processes can lead to
dephasing times $T_2$  
of some hundreds of femtoseconds for typical QDs at room 
temperature.  These values  
are of the order of the dephasing times in bulk materials and QWs, and are in 
accordance with recent room temperature measurements [\onlinecite{Borri}].

Dephasing and broadening of spectral lines in QDs 
at low temperatures ($<$ 50 K) are usually  
attributed to interactions of QD carriers with acoustic 
phonons [\onlinecite{Schmitt,Takagahara,Ota}].  In this case, the continuum of 
acoustic phonons leads to a continuous band of satellites around the 
Zero Phonon Line (ZPL).  This 
band is considered as homogeneous broadening of the formerly 
discrete QD spectral line, and the 
dephasing time is calculated from the width of the band.  
The present paper  
considers broadening and dephasing via a polar
coupling to LO phonons with fixed frequency.  
The broadening of the ZPL by LO phonons is 
interpreted as follows.  
QD carriers can virtually absorb and then emit (or emit and then absorb) 
LO phonons.  This second order 
process does not change the final energy of the 
carrier, but it does change the phase of the carrier wave function.  
This change of the phase, in turn, implies dephasing of 
the dipole for the  considered optical transition, 
and accordingly a broadening of the spectral line [\onlinecite{Osadko}].  

The scattering 
against
LO phonons gives rise to fluctuations $\Delta E(t)$ in
the transition energy  between the  electron and hole QD levels:
$E(t)=E^{(0)}+ E_s+\Delta E(t)$, where 
$E^{(0)}$ is the unperturbed transition energy,
and $E_s$ is the energy-shift due to 
the LO phonons.  For simplicity, we ignore the Coulombic
electron-hole interaction.
In a classical treatment the line structure function 
of the optical transition is given by the
Fourier transform of the correlation function
$\psi_f (\tau)={\overline{f(t+\tau)f^*(t)}}$ for the oscillator $f(t)$,
(here the overline indicates an average over fluctuations)
\begin{equation}
f(t)\propto\exp\Big\{-(i/\hbar)\big[(E^{(0)}+ E_s)t+
\int_0^tdt_1\Delta E(t_1)\big]\Big\}\;.
\label{psi}
\end{equation}
The correlation function $\psi_f(\tau)$ can be expressed
as [\onlinecite{Agrawal}]
\begin{eqnarray}
\psi_f(\tau) &\propto& \exp[-i(E^{(0)}+ E_s)(\tau/\hbar)]
\nonumber\\
&\times&\exp\left[-{1\over 2\hbar^2}\int_0^\tau dt_1 \int_0^\tau dt_2
{\overline{\Delta E(t_1) \Delta E(t_2)}}\right]\;.
\label{psif}
\end{eqnarray}
In this work we must use
the quantum mechanical analog of Eq.\,(\ref{psi}): the fluctuations 
in the transition energy are 
due to interactions with optical phonons, which
obey Bose statistics.  Thus, instead of $\psi_f(\tau)$ we 
consider [\onlinecite{Mahan}]
\begin{equation}
A(\tau)=\left\langle {\hat T} \exp\Big\{-(i/\hbar)\big[ E^{(0)}\tau +
\int_t^{t+\tau} dt' {\hat H}(t')\big]\Big\}\right\rangle,
\label{adef}
\end{equation}
where ${\hat T}$ is the time-ordering operator
and ${\hat H}(t)$ is the effective interaction Hamiltonian 
describing the fluctuations in 
the transition energy. $\langle \cdots \rangle$ indicates
an average over the phonon ensemble.
The result of our calculations, to be described below, is that
the lineshape function, which is just
the Fourier transform of (\ref{adef}),
can be
expressed in a compact form:
\begin{eqnarray}
A(\omega)&=&\int d\tau e^{i(\hbar\omega-E^{(0)}-E_s) 
(\tau/\hbar)}
\nonumber\\
&\times&\exp\Big\{- {\sigma^2\over 2\gamma_{\scriptscriptstyle LO}}
[|\tau|  + {1\over 2\gamma_{\scriptscriptstyle LO} }
(e^{-2\gamma_{\scriptscriptstyle LO}|\tau|}-1)]\Big\}\;.
\label{Afinal}
\end{eqnarray}
The shift $E_s$
and the parameter $\hbar\sigma$,
which has the meaning of average quadratic
fluctuation of the spectral line, depend on temperature, 
phonon interaction
mechanism,
and quantum dot geometry, and explicit expressions for them are given below.  
$\tau_{\scriptscriptstyle LO}=1/2\gamma_{\scriptscriptstyle LO}$ is the life-time of optical phonons,
which is finite because of various
interaction mechanisms (such as phonon-phonon scattering or
boundary scattering), and it enters our theory as a parameter 
which must be calculated separately, or extracted
from experiment. 
If $\sigma\ll 2\gamma_{\scriptscriptstyle LO}$, 
$A(\omega)$ has a Lorentzian line form with
FWHM $\Gamma=\sigma^2/\gamma_{\scriptscriptstyle LO}$, while if 
$\sigma\gg 2\gamma_{\scriptscriptstyle LO}$ 
Eq.\,(\ref{Afinal}) gives rise to a Gaussian  line with
width $\Gamma = 2\sqrt{2\ln{2}} \; \sigma$.

In order to arrive at Eq.\,(\ref{Afinal}) the following steps were needed.
First, we derive an effective Hamiltonian describing the interaction
between quantum dot charge carriers and optical phonons.  Next,  
the time-ordered expectation value (\ref{adef}) must
be evaluated.  Finally, the developed formalism will be
applied to a concrete model of a quantum dot, and the temperature dependence
of $\Gamma$ is computed. 

The carrier-phonon interaction is ($x=e$ for electrons, $x=h$
for holes) 
\begin{equation}
{\hat U}^x({\bf r},t) = \sum_{\bf q} C^x_q
\left[a_{\bf q}(t)e^{i{\bf q}\cdot{\bf r}}
+ {\rm h.c.}\right]\;,
\label{U}
\end{equation}
where  $ a_{\bf q}$ is the annihilation operator for the LO phonon 
with the wave vector ${\bf q}$.  For polar carrier--phonon coupling
$C^e_q=-C^h_q$.
The finite life-time of the optical phonons gives
rise to the following expression for the phonon Green function 
[\onlinecite{Mahan}]:
\begin{eqnarray}
D({\bf q},\tau)&\equiv&  \langle {\hat T} \{
[ a_{\bf q}(\tau) + a_{\bf q}^\dagger (\tau) ]
[ a_{\bf q}(0) + a_{\bf q}^\dagger (0) ] \} \rangle \nonumber\\
&=& e^{-\gamma_{\scriptscriptstyle LO}|\tau|}\left[({\bar n}+1)
e^{-i\omega_{\scriptscriptstyle LO} |\tau|}
+{\bar n} e^{i\omega_{\scriptscriptstyle LO} |\tau|}\right]
\end{eqnarray}
where
${\bar n}=1/[\exp(\hbar\omega_{\scriptscriptstyle LO}/k_B T)-1]$, and
$\omega_{\scriptscriptstyle LO}$ is the LO phonon frequency.
Assuming that the phonon energy $\hbar\omega_{\scriptscriptstyle LO}$
is much less than the energy separation between the QD levels, an
effective Hamiltonian ${\hat H}(t)$
can be derived
by neglecting {\it real} transitions beween the
levels [\onlinecite{Schmitt}],
and applying perturbation theory in the carrier--phonon interaction.
Up to second order one finds
\begin{eqnarray}
{\hat H}^{(1)}(t)&=& \sum_{\bf q} \left [
f({\bf q}) a_{\bf q}(t) + f^*({\bf q}) a_{\bf q}^\dagger(t) \right]\;,
\label{E1}\\
{\hat H}^{(2)}(t)&=&-\sum_{\bf qq'}
\Big[a_{\bf q}(t) a_{\bf q'}(t) F({\bf q},-{\bf q'})
\nonumber\\
&\quad&+a_{\bf q}(t) a_{\bf q'}^\dagger (t) F({\bf q},{\bf q'})
+a_{\bf q}^\dagger (t) a_{\bf q'}(t) F(-{\bf q},-{\bf q'})\nonumber\\
&\quad&+a_{\bf q}^\dagger(t) a_{\bf q'}^\dagger (t) 
F(-{\bf q},{\bf q'})\Big]\;.
\label{Omega2expl}
\end{eqnarray}
Here
\begin{eqnarray}
f({\bf q})&=&
C^e_q M^{e1}_{\bf q} + C^h_q M^{h1}_{\bf q} \;,\\
 F({\bf q},{\bf q'})&=&|C_q C_{q'}|
 \sum_{p >1} 
\left[{M^{ep}_{\bf q}{M^{ep}_{\bf q'}}^*\over E^0_{ep}-E^0_{e1}}+
{M^{hp}_{\bf q}{M^{hp}_{\bf q'}}^*\over E^0_{hp}-E^0_{h1}}\right].
\label{F(qq)}
\end{eqnarray} 
$E^0_{ep} (E^0_{hp})$ is the unperturbed electron (hole) energy
with the wave function $\psi_{ep}$ ($\psi_{hp}$), which
enters via the matrix element								
\begin{equation}
M^{xp}_{\bf q}
=\int d{\bf r} \,\psi_{xp}^*({\bf r})
\exp(i{\bf q}\cdot{\bf r})
\psi_{x1}({\bf r})\;.
\end{equation}
We consider broadening of the transition
between electron and hole ground states ($p=1$), so that
$E^{(0)}=E_g+ E^0_{e1}+E^0_{p1}$, where $E_g$ is
the energy gap of the QD material.

The first order term Eq.\,(\ref{E1}) is just the standard Huang--Rhys
Hamiltonian [\onlinecite{Schmitt}], while the 
second order term (\ref{Omega2expl}) is
an extension of this theory.
A similar second order term has earlier been shown to lead
to broadening of the ZPL in  case of impurities in doped
crystals [\onlinecite{Osadko}].  Below we show that the Hamiltonian
(\ref{E1}) -- (\ref{Omega2expl}), which only involves
virtual transitions between the QD levels,
leads to a broadening of the ZPL in QD's in accordance
with experiments [\onlinecite{Borri}].  
The existence of the nearby levels is essential
for obtaining the broadening. The effects of 
mixing of the QD levels due to transitions
between the levels on intensities of phonon
satellites have recently been considered in [\onlinecite{Fomin}].

The evaluation of (\ref{adef}) with 
the effective Hamiltonian (\ref{E1}) -- (\ref{Omega2expl}).
can be carried out exactly using the cumulant technique
[\onlinecite{Mahanagain}].  
The result is (note
that cross-terms vanish because they involve unequal number of phonons)
\begin{eqnarray}
A(\tau)&=&\exp\Big\{-i[E^{(0)}+
\sum_{j=1,2}\langle{\hat H}^{(j)}\rangle](\tau/\hbar)  
\label{Mahanformula}\\
&\quad& - {1\over 2\hbar^2} \int_0^\tau dt_1\int_0^\tau dt_2
\sum_{j=1,2}
\langle {\hat T}\Delta{\hat H}^{(j)}(t_1)
\Delta{\hat H}^{(j)}(t_2)\rangle\Big\}\;,
\nonumber
\end{eqnarray}
where
$\Delta{\hat H}^{(j)}(t)={\hat H}^{(j)}(t)-\langle {\hat H}^{(j)} \rangle$.
It is interesting to note the {\it formal} similarity with the classical result
Eq.\,(\ref{psif}).
The phonon averages can be
carried out with the help of Wick's theorem, 
which stipulates that the various operator
averages implicit in (\ref{Mahanformula})
are given as the sum of all pairwise averages.  
The time-ordered expectation value due to ${\hat H}^{(1)}$ is just
$\int_0^t dt_1\int_0^t dt_2\sum_{\bf q}|f({\bf q})|^2
D({\bf q},t_1-t_2)$;
this expression is well-known, say, from studies of phonon-assisted
tunneling in resonant-tunneling diodes [\onlinecite{Ned}]. In our case
this term gives the Huang--Rhys shift 
$E_s^{(1)}=-S \hbar\omega_{\scriptscriptstyle LO}$ of the spectral
line, where
$S=\sum_{\bf q} |f({\bf q})/(\hbar\omega_{\scriptscriptstyle LO})|^2$
is the Huang--Rhys parameter [\onlinecite{Schmitt,Mahan}], 
as well as the
{\it satellites} at energies $\pm n\hbar\omega_{\scriptscriptstyle LO}$ 
from the main
line.  For the present purposes it is important to realize
that the first-order term ${\hat H}^{(1)}$ cannot lead to
broadening of the ZPL if 
$\gamma_{\scriptscriptstyle LO}/\omega_{\scriptscriptstyle LO}\ll 1$, 
see also Ref.[\onlinecite{equal}].  A straightforward calculation shows
that the contribution due to ${\hat H}^{(2)}$ is obtained by a replacement
$|f({\bf q})|^2D(t_1-t_2) 
\to F({\bf q},{\bf q'})[F({\bf q'},{\bf q})+
F(-{\bf q},-{\bf q'})]D({\bf q},t_1-t_2)D({\bf q'},t_1-t_2)$.
This expression generates two
kinds of terms. (i) Terms proportional to 
$\exp[\pm 2i\omega_{\scriptscriptstyle LO}(t_1-t_2)]$;
together with $\langle {\hat H}^{(2)}\rangle$ in (\ref{Mahanformula})
they give the spectral shift in the second-order theory.  They
are also
relevant for the two--phonon satellites 
at $\pm n 2\hbar\omega_{\scriptscriptstyle LO}$ of the main line, 
which however is not our focus here.
(ii) Terms  without these fast time-oscillations
which
determine the zero-phonon linewidth.
Collecting all the terms describing the spectral shift
and broadening, we find 
\begin{eqnarray}
E^{(2)}_s&=&-(2{\bar n}+1)
(S_1+S_2)\hbar\omega_{\scriptscriptstyle LO}\;,
\label{Eshift}\\
\sigma&=& \sqrt{4{\bar n}({\bar n}+1)S_2} \omega_{\scriptscriptstyle LO}\;,
\label{sigma2}
\end{eqnarray}
where
\begin{eqnarray}
S_1&=&\sum_{\bf q} F({\bf q},{\bf q})/(\hbar\omega_{\scriptscriptstyle LO})\;,
\\
S_2&=&
\sum_{\bf q q'}
F({\bf q},{\bf q'})[F({\bf q'},{\bf q})+F(-{\bf q},-{\bf q'})]/
2(\hbar\omega_{\scriptscriptstyle LO})^2\;.
\end{eqnarray}
The second-order shift (\ref{Eshift}) together with the first-order
shift $E^{(1)}_s$ 
gives the total shift $E_s$ of the ZPL in (\ref{Afinal}).
The parameters $S_1$ and $S_2$ are extensions 
of the parameter $S$ of the Huang--Rhys theory.  Our
numerical calculations
show that $S_1,S_2\ll 1$, and also $S_1\gg S_2$.
Eq.\,(\ref{sigma2}) together with
(\ref{Afinal}) allows an analysis of  the spectral
lineshape, and, in particular, its width as a function of
temperature.

\begin{figure}[t]
\epsfxsize=7.5cm\epsfbox{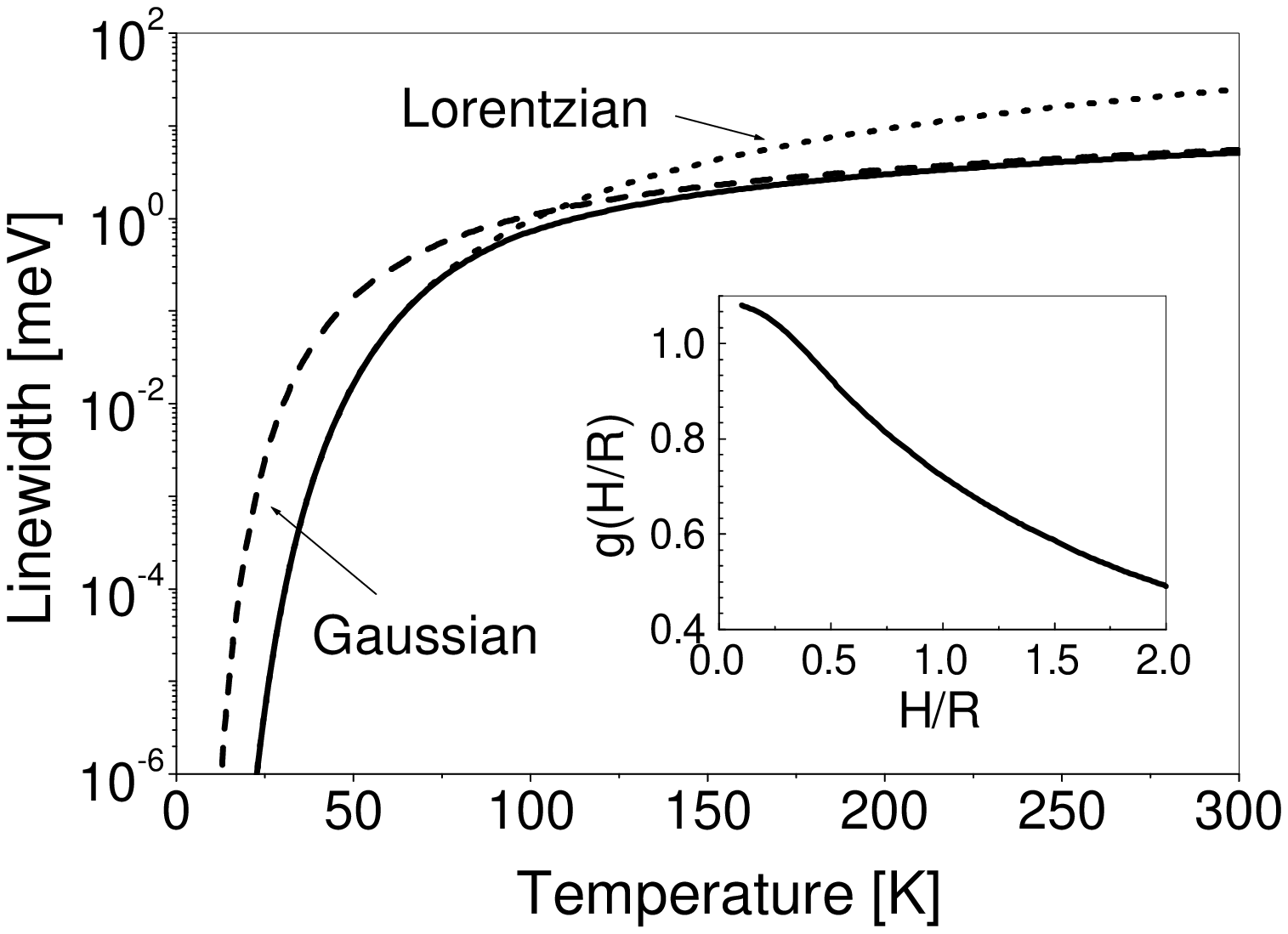}
\caption{The line-width as a function of temperature.
The following parameters were used: $\epsilon_\infty=10.9$,  
$\epsilon=12.9$, 
$m_e+m_h=0.35$, $R=H=5$ nm.  The solid curve is the result
of a full numerical calculation, while the Lorentzian and Gaussian
approximations are given as dotted and dashed lines,
respectively.  The inset shows the function $g(R/H)$.}  
\end{figure}

As a practical application, we consider
a cylindrical quantum dot of
radius $R$ and  height $H$, with infinite
confining potential.  
In the calculation of the average fluctuation  $\sigma$,
we  include in Eq.\,(\ref{F(qq)})
only the leading terms in the sum over the states $p$ .
If  $2R > H$, two states 
have the same energy, and are the closest to 
the ground state. 
Taking into account this two-fold degeneracy, we  get
\begin{equation}
\hbar\sigma = \sqrt{{\overline n}({\overline n}+1)} 
{\hbar\omega_{\scriptscriptstyle LO}
\over\Delta{\tilde E}}
g(H/R) {K\over R}
\label{sigma}
\end{equation}
where $\Delta{\tilde E}=\hbar^2/(2(m_e+m_h)R^2)[\alpha_{11}^2-
\alpha_{10}^2]$.
$\alpha_{lm}$ is the $l$-th root of the Bessel function of the order $m$.
The material parameter 
$K=e^2(1/\epsilon_\infty-1/\epsilon)/(\pi^2\epsilon_0)$ 
has the value 26.0 meV$\cdot$nm for GaAs.
The dimensionless function $g(H/R)$ is shown in the
inset of Fig. 1.

In the numerical calculation of the width $\Gamma$
(see Fig. 1)
we use  
$\tau_{\scriptscriptstyle LO} =1.5$ ps [\onlinecite{Ridley}]
so that 
$\hbar\gamma_{\scriptscriptstyle LO} = 0.22$ meV and 
$(\omega_{\scriptscriptstyle LO}
\tau_{\scriptscriptstyle LO})^{-1}\simeq 0.01$.  
From (\ref{sigma}) one finds that at 
$T=300$ K the average quadratic fluctuation  is
$\hbar\sigma\simeq 2.2$ meV
(corresponding to $S_2\simeq 2\cdot 10^{-3}$ in Eq.\,(\ref{sigma2})), so that the 
condition $\sigma\gg 2 \gamma_{\scriptscriptstyle LO}$
is satisfied, and consequently the spectral line has a Gaussian shape.
At lower temperatures $\sigma$ decreases and can reach
the regime $\sigma\ll 2\gamma_{\scriptscriptstyle LO}$, where the Lorentzian shape
obtains. 
At $T=$ 300 K the linewidth $\hbar\Gamma$ is 5.1 meV,
corresponding to  
$T_2 \simeq 250$ fs. 
Taking into account higher excited states in the sum (\ref{F(qq)})
in evaluating
$\sigma$  typically reduces $T_2$ by a factor of $1.5-2$. 
Thus, one can 
expect  room temperature 
dephasing times  of the order of $100 - 200$ fs.  
This is in good agreement with experimental values ($200 - 300$ fs)
[\onlinecite{Borri}].
The LO-phonon contribution to the linewidth is proportional to 
${\overline n}({\overline n}+1)$ if 
$\Gamma\ll 1/\tau_{LO}$, and
thus vanishes as the temperature
tends to zero.  For example, for $T<50$ K the calculated
values of $\hbar\Gamma$ are below 20 $\mu$eV (see Figure 1), 
which is below the linewidths
found in experiment ($\simeq 50 - 150$  $\mu$eV) 
[\onlinecite{Takagahara,Ota}].
On the other hand, the acoustic phonon related linewidth does
{\it not} vanish in this limit [\onlinecite{Takagahara}], 
and we conclude
that the LO-mechanism is dominant
only at elevated temperatures, say $T>$ 100 K. 
The present broadening mechanism 
also provides a possible way of understanding 
fast carrier relaxation  in QDs at elevated temperatures
[\onlinecite{Sosnowski}]:
due to the substantial broading of the levels their
differences no longer have to match strictly the LO-phonon
energy, thereby weakening the phonon bottleneck effect.

It is also instructive to consider the classical limit of our
calculations.  Then ${\overline n}\gg$ 1, and thus for
$\Gamma\ll 1/\tau_{\scriptscriptstyle LO}$ we find $\Gamma\propto {\overline n}^2$.
This result also follows from Eq.\,(\ref{psif}) if we treat the
phonons as classical random fields perturbing the QD carriers.
This classical result allows one to establish an analogy
with broadening in  atomic gases [\onlinecite{Demtroeder}].
There, elastic atomic collisions lead to dephasing of
optical transitions, and the homogeneous linewidth in
the Lorentzian limit is proportional to the density of the
colliding particles.  In our case the broadening of
ZPL is the result of elastic {\it second order}
interactions between phonons and the carriers, and
hence $\Gamma\propto{\overline n}^2$.  Note also that
the LO phonon lifetime $\tau_{\scriptscriptstyle LO}$ plays the same
role as the collisional time in atomic gases
[\onlinecite{Demtroeder}].

We finally consider the effect of the shape and size of the
QD on the dephasing time.  Eq.\,(\ref{sigma}) implies
that $\sigma\propto L$, where $L$ characterizes the size
of the QD.  In the Lorentz case $\Gamma\propto\sigma^2\propto L^2$,
while in the Gaussian case $\Gamma\propto\sigma\propto L$.
From Fig. 1 we can see that $\Gamma$ can be modified by a
factor 2--3 by changing the ratio $H/R$ of the cylinder.
Note also that $\sigma$ and $\Gamma$ depend on the
carrier masses, $\sigma\propto (m_e+m_h)$.
These characteristic dependencies point towards a possibility
to {\it engineer} the spectral linewidths.

In conclusion, we have developed a theory for dephasing
of an optical transition in QDs due to second-order elastic
interaction with LO-phonons.  The theory results in expressions
which are easily evaluated numerically, and can also
be used to draw qualitative conclusions
about the dependence of the dephasing time on physical parameters
defining the QD.  At room temperature, despite of an apparent
phonon bottleneck, the considered mechanism leads to dephasing
times of the order of 200 fs, i.e. of the same order as in bulk,
and also as observed in QDs. 

One of us (A.~V.~U.) thanks Ministry of Education, Science, Culture and
Sport of Japan for support during his work at TUAT.
We acknowledge D.~Birkedal, P.~Borri, and J.~Hvam for
useful comments.



\begin{thebibliography}{10}
\bibitem{Mahanagain} A discussion of models which can be solved exactly
with the linked cluster expansion can be found in 
Sect. 4.3.F of Ref. \protect\onlinecite{Mahan}.
Obviously, the full electron--phonon interaction (\protect\ref{U})
does not belong to this class.
\bibitem{equal}
We also note that 
the amplitude $f({\bf q})$ is in small, because the electron
and hole contributions tend to cancel for polar coupling.  In particular, 
$f({\bf q})\equiv 0$,
if the electron and hole wave-functions are equal, such as in an
infinitely deep well.
\end{thebibliography}
\end{document}